\documentclass[twocolumn,showpacs,preprintnumbers,amsmath,amssymb,prb,aps]{revtex4-1}
\usepackage{epsfig}
\usepackage{epstopdf}
\usepackage{graphicx}
\usepackage{dcolumn}
\usepackage{bm}
\usepackage{slashbox}
\usepackage{textcomp}
\usepackage{array}
\usepackage{subcaption}
\usepackage{booktabs}

\begin{document}

\title{Experimental and computational approaches to study the high temperature thermoelectric properties of novel topological semimetal CoSi}

\author{Shamim Sk$^{1,}$}
\altaffiliation{Electronic mail: shamimsk20@gmail.com}
\author{Nisha Shahi$^{2}$}
\author{Sudhir K. Pandey$^{3,}$}
\altaffiliation{Electronic mail: sudhir@iitmandi.ac.in}
\affiliation{$^{1}$School of Basic Sciences, Indian Institute of Technology Mandi, Kamand - 175075, India}
\affiliation{$^{2}$School of Materials Science and Technology, Indian Institute of Technology (Banaras Hindu University), Varanasi 221005, India}
\affiliation{$^{3}$School of Engineering, Indian Institute of Technology Mandi, Kamand - 175075, India}


\begin{abstract}
Here, we study the thermoelectric properties of topological semimetal CoSi in the temperature range $300-800$ K by using combined experimental and density functional theory (DFT) based methods. CoSi is synthesized using arc melting technique and the Rietveld refinement gives the lattice parameters of a = b = c = 4.445 \AA \,. The measured values of Seebeck coefficient (S) shows the non-monotonic behaviour in the studied temperature range with the value of $\sim-$81 $\mu$V/K at room temperature. The $|S|$ first increases till 560 K ($\sim-$93 $\mu$V/K) and then decreases up to 800 K ($\sim-$84 $\mu$V/K) indicating the dominating \textit{n}-type behaviour in the full temperature range. The electrical conductivity, $\sigma$ (thermal conductivity, $\kappa$) shows the monotonic decreasing (increasing) behaviour with the values of $\sim$5.2$\times 10^{5}$ (12.1 W/m-K) and $\sim$3.6$\times 10^{5}$ (14.2 W/m-K) $\Omega^{-1}m^{-1}$ at 300 K and 800 K, respectively. The $\kappa$ exhibits the temperature dependency as, $\kappa \propto T^{0.16}$. The DFT based Boltzmann transport theory is used to understand these behaviour. The multi-band electron and hole pockets appear to be mainly responsible for deciding the temperature dependent transport behaviour. Specifically, the decrease in the $|$S$|$ above 560 K and change in the slope of $\sigma$ around 450 K are due to the contribution of thermally generated charge carriers from the hole pockets. The temperature dependent relaxation time ($\tau$) is computed by comparing the experimental $\sigma$ with calculated $\sigma/\tau$ and it shows temperature dependency of $1/T^{0.35}$. Further this value of $\tau$ is used to calculate the temperature dependent electronic part of thermal conductivity ($\kappa_{e}$) and it gives fairly good match with the experiment. Present study suggests that electronic band-structure obtained from DFT provides reasonably good estimate of the transport coefficients of CoSi in the high temperature region of $300-800$ K.       
                          
\vspace{0.3cm}
Key words: Thermoelectric properties, density functional theory, Spin-orbit coupling, electronic structure, Boltzmann theory, figure of merit.

\end{abstract}

\maketitle
\section{INTRODUCTION}
Theoretical understanding of the transport properties of materials has always been a challenging task. One of the key quantity in understanding the transport properties is relaxation time ($\tau$). The theoretical estimation of $\tau$ is very difficult in any real system due to presence of various scattering mechanisms such as, electron-electron interaction, electron-phonon interaction, phonon-phonon interaction, electron-defect interaction, phonon-defect interaction etc\cite{ashcroft}. The visualization of these scattering mechanisms is really challenging at the level of theory as well as computation due to the complexity involved in the many-body interactions. Along with this, the calculation of transport properties requires dense k-point sampling demanding larger computational cost. However, recent development of high performance computers improves the situation to some extent as compared to earlier days. But, still calculating the temperature dependent transport properties are much complicated. Because, in this case calculation is needed to be done at every temperature which is time consuming and computationally expensive job. For instance, the electrical conductivity ($\sigma$) of a material can be calculated as\cite{ashcroft},

\begin{equation}
\boldsymbol{\sigma}^{(n)} = e^{2}\int\frac{d\textbf{k}}{4\pi^{3}}\tau_{n}(\varepsilon_{n}(\textbf{k}))\textbf{v}_{n}(\textbf{k})\textbf{v}_{n}(\textbf{k})\bigg(-\frac{\partial f}{\partial \varepsilon}\bigg)_{\varepsilon=\varepsilon_{n}(\textbf{k})}
\end{equation}
for $n^{th}$ band. Where, $e$ is an electronic charge, $\tau_{n}(\varepsilon_{n}(\textbf{k}))$ and $\textbf{v}_{n}(\textbf{k})$ are relaxation time and mean velocity of an electron of $n^{th}$ band with wave vector \textbf{k}, respectively. The $\varepsilon_{n}(\textbf{k})$ is an energy band and $\frac{\partial f}{\partial \varepsilon}$ is the partial energy derivative of Fermi-Dirac distribution function. The total $\sigma$ is calculated by adding all the $\sigma$'s corresponding to each band as: $\boldsymbol{\sigma} = \sum_{n}\boldsymbol{\sigma}^{(n)}$. Therefore, to calculate the temperature dependent $\sigma$, all the terms appear in Eqn. 1 should be calculated at every temperature, which is a difficult and time consuming task. However, the recent advances in computational power enables to calculate the temperature dependent transport properties up to some extent. For instance, density functional theory + dynamical mean field theory (DFT+DMFT) is used to calculate temperature dependent electronic structure and hence related transport properties\cite{dmft1,dmft2,dmft3,dutta_2019,dutta_2020}. But, the computational cost of these techniques is high as well as implementation of these techniques is not straightforward. In this scenario, one of the reliable and cheap method to understand the electronic properties is DFT proposed by Hohenberg and Kohn\cite{dft}. 

DFT calculates the ground state properties of the materials by solving the Kohn-Sham equation. One of the easy way to calculate the temperature dependent transport properties is by taking ground state of DFT and solving the semi-classical Boltzmann transport equation. But, the realization of ground state properties under DFT itself faces many challenges. For instance, in DFT many-body wave function is replaced by single-particle wave function. In addition to this, the DFT result is approximated by introducing the different exchange-correlation (XC) functionals. Among the many XC functionals developed, the local density approximation (LDA)\cite{lda} and generalized gradient approximations (GGAs)\cite{pbe} are the popular and widely used functionals in condensed matter physics. The accuracy of DFT result depends on how smartly XC functional is chosen. On the top of the fact that by taking the ground state and single-particle wave function, it will be interesting to see up to what extent DFT explains the transport properties of the materials at high temperature. Generally, Fermi-Dirac distribution function takes care of the temperature dependency of the transport properties. In this context, it should be noted that ground state structure may not be appropriate for the temperature dependent study, because band-structure itself has temperature dependence. Apart from this, the temperature dependence of band gap, scattering phenomena etc. make the study of high temperature transport properties more challenging. Keeping all these challenges, here we are interested to understand all the experimentally observed thermoelectric parameters: Seebeck coefficient (S), electrical conductivity ($\sigma$) and thermal conductivity ($\kappa$) using first-principles DFT calculations.

The study of thermoelectric (TE) materials is of great interest in recent decades as they generate electricity from waste heat\cite{akasaka,pei}. These materials are believed to hold the key for clean energy production in near future. The efficiency of TE materials are evaluated by unitless parameter called figure of merit\cite{zt}, $ZT=S^{2}\sigma T/\kappa$. Where, T is absolute temperature of the material. The $\kappa$ involves two parts: electronic thermal conductivity ($\kappa_{e}$) and lattice thermal conductivity ($\kappa_{L}$). The efficient TE materials should possess $ZT\geq1$\cite{snyder}. Therefore, good TE materials should have high power factor (PF = $S^{2}\sigma$) but low $\kappa$. However, the poor efficiency of TE materials limit their commercial use specially at high temperature region. Realization of high \textit{ZT} is really a difficult job. Because S, $\sigma$ and $\kappa_{e}$ are deeply interrelated to each other through charge carrier\cite{ashcroft,shamim_mrx}. Enhancement of $\sigma$ without affecting $\kappa_{e}$ or minimizing $\kappa_{e}$ without disturbing $\sigma$ is a terrible job as they involve linear relationship via Wiedeman-Franz law: $\kappa_{e}=L\sigma T$, \textit{L} is Lorenz number.             

Here, we have taken CoSi as a case study for understanding all the experimentally observed TE parameters using DFT at high temperature. As per the definition, CoSi comes under the category of strongly correlated electron system (SCES), because Co contains partially filled 3\textit{d} orbitals. Many previous studies show that DFT is not capable to produce the correct electronic structure for such a SCES\cite{dft_fails1,dft_fails2,dft_fails3}. But, recent study of Dutta \textit{et al.}\cite{dutta_2019} shows that correlations among Co 3\textit{d} electrons in CoSi are weak and DFT was seen to give more accurate electronic structure. Therefore, it is expected that further extension of DFT results will address the temperature dependent experimental TE properties of CoSi properly.    

CoSi with a B20 simple cubic structure of space group $P2_{1}3$ (No. 198) has been reported as a promising TE candidate due to having large PF at room temperature\cite{asanable,kim,lue,ren,li,kuo,pan,skoug,sun_2013,sun_2017,longhin,yu}. PF of CoSi is comparable with those of state-of-the-art TE materials Bi$_{2}$Te$_{3}$\cite{bite} and PbTe\cite{pbte}. But, the benefits of high PF are diminished by it's high $\kappa$, resulting in a low \textit{ZT}. CoSi comes under transition metal silicides. This class of materials have been gained great attention from last few decades due to their practical application in electronics, magnetism and thermoelectrics\cite{lange,shinoda,wernick,grigoriev,paschen,han,nakanishi,dutta_2018,dutta_2019}. Among them, CoSi has been recently marked as nontrivial topological semimetal\cite{tang,takane}. Unlike other semimetals, band-structure of CoSi holds linear touching points of bands at $\Gamma$ and \textit{R} points in the vicinity of Fermi level. These points are usually called as nodes or nodal points. This unusual features of band-structure around the Fermi level motivates us to choose CoSi as a case study for understanding all the TE properties at high temperature using combined experimental and theoretical tools. The experimental TE properties of parent CoSi is reported by many groups as mentioned earlier. Specifically, the S of this compound is reported in the range of $\sim-50$ to $-90$ $\mu$V/K at 300 K in the various literatures\cite{kim,lue,ren,li,pan,sun_2013,sun_2017}. The variation in S values at 300 K suggests the off-stoichiometry of the compound. This off-stoichiometry generally comes during the sample synthesis by error in weighing the starting materials, evaporation of low melting element, inhomogeneous mixing etc. By considering all these factors and taking dense data points at high temperature, the proper understanding of all the TE properties of CoSi is lacking from the literatures. This gives us the motivation to study the same in this direction.        

In this work, we present the theoretical analysis of experimentally observed TE properties of CoSi in the temperature range $300-800$ K. The CoSi has been synthesized using arc melting technique and the lattice parameters of a = b = c = 4.445 \AA \, are obtained from the Rietveld refinement. The S of CoSi is observed as $\sim-$81 $\mu$V/K at 300 K. The $|S|$ increases first till 560 K with the value of $\sim-$93 $\mu$V/K and then decreases up to 800 K ($\sim-$84 $\mu$V/K). Measurement of S indicates the dominating \textit{n}-type behaviour of the compound in the full temperature range. The values of $\sigma$ are observed as $\sim$5.2$\times 10^{5}$ and $\sim$3.6$\times 10^{5}$ $\Omega^{-1}m^{-1}$ at 300 K and 800 K, respectively. The monotonic decrement of $\sigma$ with temperature is noticed in the full temperature window, whereas temperature dependent $\kappa$ are found to increase in the studied temperature range with the values of 12.1 and 14.2 W/m-K at 300 K and 800 K, respectively. The behaviour of all these temperature dependent transport properties are understood by combined DFT and Boltzmann theory. The multi-band electron and hole pockets are found to give quite good explanation of temperature dependent transport coefficient. Temperature dependence of $\tau$ is calculated, which shows the temperature dependency of $1/T^{0.35}$. \textit{ZT} and efficiency are calculated using the experimental transport coefficient. Maximum \textit{ZT} is found to be $\sim$0.15 at 650 K. A systematic understanding of all the TE parameters of CoSi are made using combined experimental and first-principles DFT based methods through present study.

\begin{figure}
\includegraphics[width=0.97\linewidth, height=7.1cm]{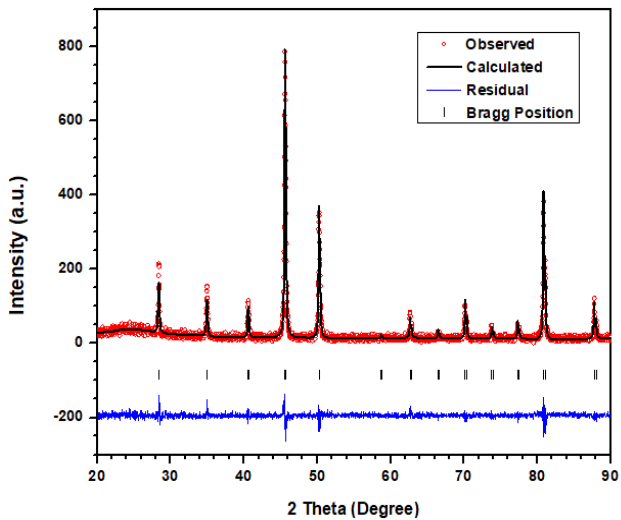} 
\caption{\small{Room temperature X-ray diffraction of CoSi.}}
\end{figure}

\section{EXPERIMENTAL AND COMPUTATIONAL DETAILS}
The polycrystalline ingots of CoSi was synthesized using arc melting technique. The high purity element of Co (99.99$\%$) and Si (99.99$\%$) were taken as the starting materials. The ingots of CoSi were obtained by melting the appropriate amounts of Co and Si in a vacuum arc furnace. The room temperature x-ray diffraction (XRD) with CuK$\alpha$ radiation (1.5406 \AA) in the 2$\theta$ range of $20-90^{o}$ has been performed. The B20 cubic crystal structure was confirmed from the Rietveld refinement method through the validation of XRD pattern as shown in Fig. 1. The lattice parameters of a = b = c = 4.445 \AA \, were obtained from the refinement. The refined Wyckoff positions of Co and Si atoms are 4a (0.160, 0.160, 0.160) and 4a (0.855, 0.855, 0.855), respectively.

Measurement of S and $\kappa$ were carried out using home-made experimental setup\cite{shamim_instrument}. The sample with dimension 6 mm $\times$ 4mm $\times$ 2mm (length $\times$ width $\times$ thickness) was used for the measurement. Resistivity was also measured using the home-made setup\cite{saurabh_resistivity}.

The ground state electronic structure calculations are carried out within density functional theory (DFT)\cite{dft} using full-potential augmented plane wave (FP-LAPW) method as implemented in WIEN2k code\cite{wien2k}. The calculation is done in the presence of spin-orbit coupling (SOC). Local density approximation (LDA)\cite{lda} is used as an XC functional. The \textit{muffin-tin} sphere radii (R$_{MT}$) of 2.34 and 1.73 Bohr are chosen for Co and Si, respectively. The convergence criteria for the calculation of ground state energy is set as 0.1 mRy/cell. The electronic structure related transport properties are calculated using BolzTraP package based on Boltzmann semi-classical transport theory\cite{boltztrap}. A heavy k-mesh of size 40 $\times$ 40 $\times$ 40 is used in the ground state electronic structure calculations to aid the calculation of transport properties.

\begin{figure*}
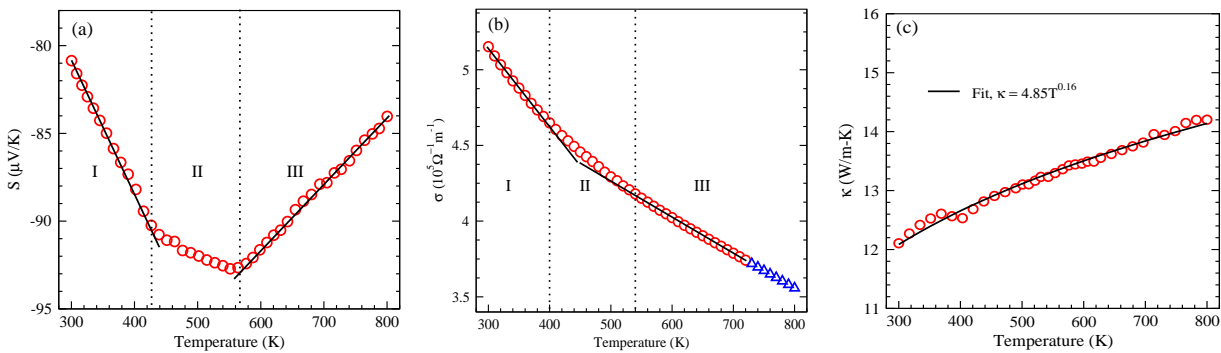
 
\begin{subfigure}{0.3\textwidth}
\includegraphics[width=0.98\linewidth, height=4.6cm]{fig2a_seebeck_exp.eps} 
\end{subfigure}
\begin{subfigure}{0.3\textwidth}
\includegraphics[width=0.95\linewidth, height=4.6cm]{fig2b_sigma_exp.eps}
\end{subfigure} 
\begin{subfigure}{0.3\textwidth}
\includegraphics[width=0.95\linewidth, height=4.6cm]{fig2c_kappa_exp.eps}
\end{subfigure} 
\caption{\small{Temperature dependence of (a) Seebeck coefficient, (b) electrical conductivity (Triange symbols denote extrapolated data) and (c) thermal conductivity.}}
\label{fig:image2}
\end{figure*}

\section{RESULTS AND DISCUSSION}        
\subsection{EXPERIMENTAL TRANSPORT PROPERTIES}
Fig. 2(a) shows the seebeck coefficient (S) of CoSi measured in the temperature range $300-800$ K. The S at 300 K is found to be $\sim-$81 $\mu$V/K. The increment/decrement trend of S is divided into three parts I, II and III as shown in figure. From the figure it is noticed that in region I, with increase in temperature the magnitude of S increases till $\sim$425 K with the value of $\sim-$90 $\mu$V/K. After 425 K in region II, the increment rate of magnitude of S decreases upto $\sim$560 K with the value of $\sim-$93 $\mu$V/K. After 560 K in region III, the magnitude of S decreases up to the highest temperature. At 800 K the S is measured as $\sim-$84 $\mu$V/K. The temperature dependent behaviour of S is similar to the previous reported works\cite{severin,solomkin}. The non-linear trend of S can be understood with the help of following equation:\cite{ashcroft}
\begin{equation}
S = \bigg(\frac{4\pi^{2}k_{B}^{2}}{3eh^{2}}\bigg)\bigg(\frac{\pi}{3n}\bigg)^{2/3}m^{*}T,
\end{equation}
where $n$ is the carrier density, $m^{*}$ is the  effective mass of the carrier and $T$ is the absolute temperature. All the other parameters are constants and have their usual meaning. From above equation, it is clear that the value of S depends on $n$, $m^{*}$ and $T$. Among them $n$ and $T$ are positive quantity. Then sign of $S$ is solely decided by $m^{*}$ only. The electrons and holes both contribute in total S, where electrons give negative S and holes yield positive S. The observed S in the full temperature range is negative here, which signifies the dominating \textit{n}-type behaviour of the compound. As the temperature increases from 300 K to $\sim$560 K, the contribution rate of electron in S increases resulting in increase of magnitude of S. After 560 K, still electrons are dominating the behaviour of S, but rate of contribution of electrons decreases with increase in temperature. Therefore, the decrement of magnitude of S at high temperature should be related to rapid increment of $n$ and/or decrement of $m^{*}$ as Eqn. 2 says. The electronic band-features is expecting to give the proper explanation of this behaviour of S which is discussed later.

Temperature dependence of electrical conductivity ($\sigma$) is exhibited in Fig. 2(b). The $\sigma$ were taken in the temperature ranage $300-720$ K. From 730 to 800 K, $\sigma$ are extrapolated (indicated by blue triangle symbols in the figure) in order to calculate \textit{ZT} as the S and $\kappa$ are already measured till 800 K. This extrapolation is expected not to disturb the value of \textit{ZT} as the linear behaviour of $\sigma$ is reported previously in this temperature range\cite{li,severin}. The values of $\sigma$ are found to be $\sim$5.2$\times 10^{5}$ and $\sim$3.6$\times 10^{5}$ $\Omega^{-1}m^{-1}$ at 300 K and 800 K, respectively. The $\sigma$ gradually decreases with increasing temperature in the full temperature window, consistent with the earlier works\cite{li,ren,severin}. But, the decrement rates are different in regions I, II and III as shown in figure as similar response was seen in the case of S. The decrement of $\sigma$ with temperature can be understood by the following well known relation: $\sigma=\dfrac{ne^{2}\tau}{m^{*}}$. Here, it is important to note that with increase in temperature $n$ always increases, while $\tau$ always decreases. Among $n$ and $\tau$, the dominating quantity generally gives the temperature dependent behaviour of $\sigma$. Typically, $\sigma$ for metals decreases with increase in temperature. This is because of increment of scattering at high temperature which is responsible for the reduction of $\tau$ and hence decrement of $\sigma$\cite{alekseeva}. But, in the case of semiconductors increment of $n$ is more dominant than decrement of $\tau$ results in increment of $\sigma$ with temperature. In the present case CoSi is known as semimetal and hence it's temperature dependent $\sigma$ is expecting to lie between the temperature dependent behaviour of $\sigma$ of metals and semiconductors. In this study, the decreasing nature of $\sigma$ with temperature suggesting that decrement of $\tau$ is dominant quantity over increment of $n$. However, the electronic band-structure is expecting to provide the proper explanation of observed experimental $\sigma$.     

The thermal conductivity ($\kappa$) is measured in the temperature range $300-800$ K as displayed in Fig. 2(c). The value of $\kappa$ at 300 K is found to be $\sim$12.1 W/m-K, while this value is observed as $\sim$14.2 W/m-K at 800 K. Figure shows that the $\kappa$ is increasing monotonically with increase in temperature, consistent with a previous work\cite{kim}. The values of $\kappa$ are quite high as compared to commercially used TE materials. We know that the total thermal conductivity comes from two parts: electronic contribution $\kappa_{e}$ and lattice contribution $\kappa_{L}$: $\kappa=\kappa_{e}+\kappa_{L}$. Where, $\kappa_{e}$ is directly proportional to the $\sigma$ through Wiedeman-Franz law. The $\kappa$ shows the temperature dependency as, $\kappa \propto T^{0.16}$ as shown in the same figure.

\subsection{ELECTRONIC STRUCTURE}
In order to understand the experimentally observed electronic transport properties (S, $\sigma$ and $\kappa_{e}$), we have carried out the electronic structure calculations. Fig. 3(a) shows the calculated electronic dispersion of CoSi with SOC along the high symmetry directions $\Gamma-X-M-\Gamma-R-M-X$ in the first Brillouin zone. The horizontal dashed line corresponding to 0 energy denotes the Fermi level ($E_{F}$). From the figure it is clear that few bands cross the $E_{F}$ at either side of $\Gamma$ and $R$-points. In general, this type of little mixing of bands around the $E_{F}$ predicts the semimetal like behaviour of the compound, which is consistent with the other works\cite{pan,samatham}.

\begin{figure*}
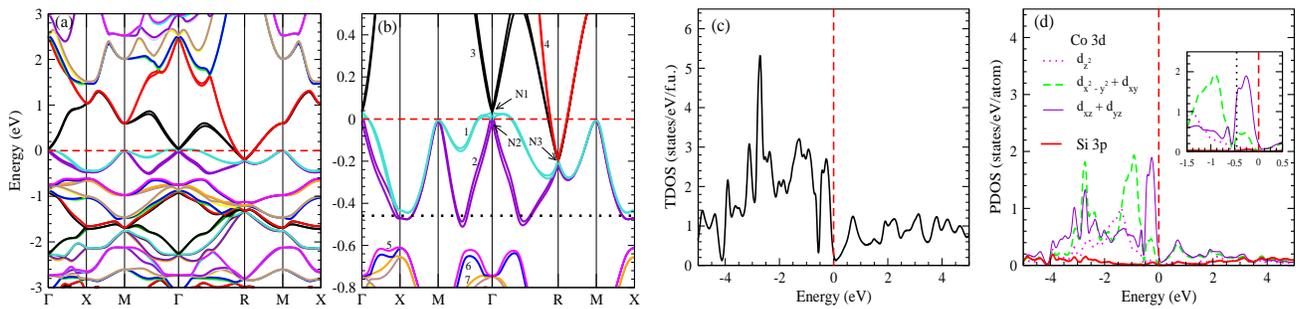
 
\begin{subfigure}{0.48\textwidth}
\includegraphics[width=0.98\linewidth, height=4.0cm]{fig3ab_bands.eps} 
\end{subfigure}
\begin{subfigure}{0.48\textwidth}
\includegraphics[width=0.98\linewidth, height=4.0cm]{fig3cd_dos.eps}
\end{subfigure} 
\caption{\small{(a) The calculated band-structure. Red dashed line indicates the Fermi level. (b) Enlargement of band-structure. N1, N2 and N3 are nodal points. Black dotted line indicates where transport properties are calculated. (c) Total density of states (TDOS). (d) Partial density of states (PDOS). Inset shows the enlargement of PDOS around the $E_{F}$.}}
\label{fig:image2}
\end{figure*}

Electronic dispersion plays a main role for understanding the electronic transport properties of any material. Because, the key input for calculating electronic transport properties is electronic dispersion. The band features around $E_{F}$ generally gives the transport properties. Therefore, for the better understanding, enlargement of band-structure are drawn in the short energy window of $-$0.8 to 0.5 eV as shown in Fig. 3(b). Four bands indicated by numbers 1, 2, 3, 4 around the $E_{F}$ are mainly expecting to contribute in the transport properties. From the figure it is clear that by the inclusion of SOC each band is splited into two. Three nodal points marked as N1, N2 and N3 are observed in the figure. Among them two nodal points appear at $\Gamma$ (just below and above the $E_{F}$) and another one at $R$-point (below $E_{F}$ at $\sim-0.2$ eV). Corresponding to these nodal points CoSi is reported as nontrivial topological semimetal\cite{tang,takane}. From the figure, we can also notice that around $E_{F}$, the hole pockets at $\Gamma$ point and electron pockets at \textit{R} point are mainly contributing in the electronic states. Using the electronic dispersion discussed here, different transport coefficient are calculated which are described in the next sub-section and compared with the experiment. 

Here one can define the effective mass ($m^{*}$) which has the important role in understanding the magnitude as well as sign of S. Under parabolic approximation, $m^{*}$ is expressed as: $m^{*}=\hbar^{2}/(d^{2}E/dk^{2})$. This formula implies that flat band has the higher $m^{*}$ as compared to the curved band. Therefore, flat band will have the larger S as compared to curved band, since S is proportional to $m^{*}$ (Eqn. 2). Convex band (hole pocket) corresponds to positive S, while concave band (electron pocket) gives the negative S. This will be more clear once we discuss the calculated S in the next sub-section.                

Fig. 3(c) displays the calculated total density of states (TDOS) of CoSi. TDOS shows the minimum DOS of $\sim$0.3 states/eV/f.u. at $E_{F}$. According to Mott\cite{mott}, this minimum state at $E_{F}$ signifies the presence of pseudogap which pushes the compound in the class of semimetals. In accordance with this, the semimetallic behaviour of CoSi with pseudogap is reported previously\cite{pan,samatham}. To see the contribution in TDOS from different atomic orbitals of CoSi, we have calculated partial density of states (PDOS) as shown in Fig. 3(d). The dominant contribution in PDOS mainly comes from Co 3\textit{d} orbitals with negligibly small contribution from Si 3\textit{p} orbitals. In the low lying energy range of $\sim-$0.5 to 0 eV, the main contribution in DOS comes from $d_{xz}+d_{yz}$ orbitals with small contribution from $d_{x^{2}-y^{2}}+d_{xy}$ orbitals. Above $E_{F}$, the $d_{xz}+d_{yz}$ and $d_{x^{2}-y^{2}}+d_{xy}$ orbitals are almost equally contributed in DOS. From the figure it is observed that the contribution of Si 3\textit{p} orbitals in DOS is quite low expecting to give a very small contribution in transport properties. Hence, it is evident from the figure that $d_{xz}+d_{yz}$ orbitals are mainly responsible for the transport properties of CoSi.

\begin{figure*}
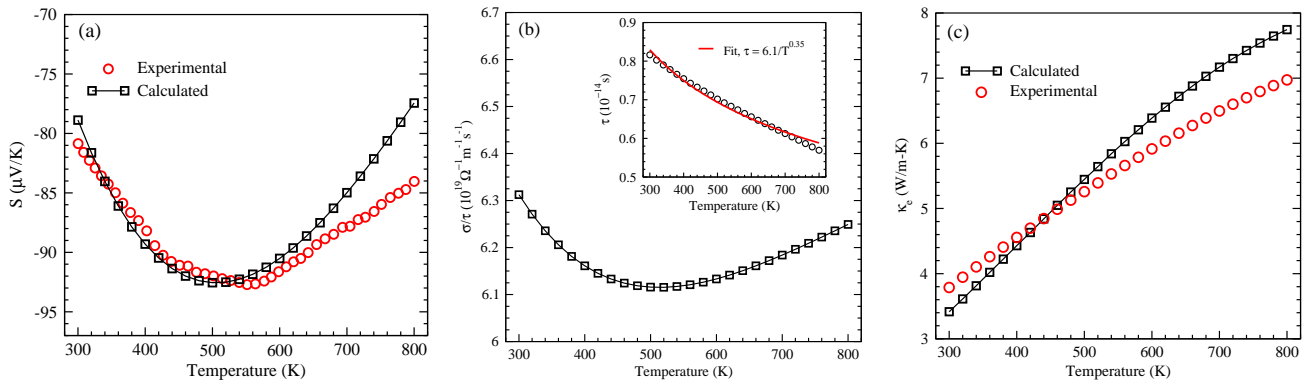
 
\begin{subfigure}{0.32\textwidth}
\includegraphics[width=0.98\linewidth, height=5.0cm]{fig4a_seebeck_exp_cal.eps} 
\end{subfigure}
\begin{subfigure}{0.32\textwidth}
\includegraphics[width=0.95\linewidth, height=5.0cm]{fig4b_sigma_cal.eps}
\end{subfigure} 
\begin{subfigure}{0.32\textwidth}
\includegraphics[width=0.95\linewidth, height=5.0cm]{fig4c_ke_cal.eps}
\end{subfigure} 
\caption{\small{(a) Comparison of experimental and calculated values of Seebeck coefficient. (b) Calculated electrical conductivity per relaxation time ($\tau$). Inset shows $\tau$ as a function of temperature extracted by comparing the calculated $\sigma /\tau$ with experimental $\sigma$. Solid red line of inset indicates the fitting between $\tau$ and  temperature. (c) Experimental (estimated) and calculated electronic part of thermal conductivity.}}
\label{fig:image2}
\end{figure*}

\subsection{CALCULATED TRANSPORT PROPERTIES}
In this section, the Seebeck coefficient (S), electrical conductivity ($\sigma$) and electronic part of thermal conductivity ($\kappa_{e}$) are described which are calculated using semi-classical Boltzmann theory as implemented in BoltzTraP code\cite{boltztrap}. Then experimentally observed transport properties are understood through calculated values. At room temperature the calculated value of S is $\sim$44 $\mu$V/K at $E_{F}$. This indicates the dominating \textit{p}-type behaviour of the compound. We can see from the band-structure of Fig. 3(b) that the hole and electron pockets are appeared at $\Gamma$ and \textit{R} points, respectively. With respect to $E_{F}$, the energy gap of hole pockets is smaller than electron pockets. Therefore, holes from hole pocket will be easily excited than the electrons from electron pocket. Apart from this, among the two bands (band 1 splits into two) making hole pocket at $\Gamma$ one is flat and which has the higher $m^{*}$. That's why dominating contribution in S comes from holes at $E_{F}$. However, this positive value of S at $E_{F}$ is much far away from the experimental negative value of $\sim-$81 $\mu$V/K at the same temperature. At this point it is important to note that the calculation is done on single crystalline stoichiometric compound. But, the experimentally measured S is negative for CoSi. This observed difference can be attributed to the off-stoichiometry in the experimental sample. Normally, there is a chance of off-stoichiometry in the synthesized samples as discussed in introduction. This factor can be taken into account in the calculation by shifting the position of chemical potential ($\mu$). We found at $\mu$ $\approx$ $-$459 meV, the calculated value of S is $\sim-$79 $\mu$V/K at 300 K, which closely matches with the experimental value. Using this $\mu$, the values of S in the temperature range $300-800$ are calculated and compared with experiment as shown in Fig. 4(a). Calculated S gives the good match with experimental S but shows slight deviation at higher temperature region ($>$600 K). From the inset of Fig. 3(d), it is clear that $d_{xz}+d_{yz}$ orbitals are mainly contributing in DOS at $\sim-$0.459 meV and expecting to contribute in S. This $\mu$ value corresponds to the electrons deficiency of 0.88/f.u. 

At $\mu$ $\approx$ $-$459 meV (marked by black dotted line), three electron pockets (one at the vicinity of $X$-point and another two are in $\Gamma-M$ $\&$ $\Gamma-R$ directions) are observed in Fig. 3(b). The presence of electron packets (which give negative S) at $\sim$ $-$459 meV supports the negative value of S in the calculation using the same dispersion. As the temperature increases, the $|S|$ is expected to be increased according to Eqn. 2 and which is exactly seen in Fig. 4(a). But after $\sim$425 K, the increment rate of $|S|$ decreases up to $\sim$560 K. After 560 K, the $|S|$ decreases. This non-monotonic behaviour of S is because of contribution of hole pockets (which give positive S) starts at higher temperature. These hole pockets are around the vicinity of $X$-point, in $\Gamma-M$ and $\Gamma-R$ directions at $\sim-$0.6 eV (below $\sim$140 meV from the energy level where S is calculated) of Fig. 3(b). Specifically after 425 K, the increment rate of $|S|$ decreases due to the contribution of hole pockets of band 5. Around the vicinity of $X$-point at $\sim-$0.6 eV, band 5 is flat which will have the larger $m^{*}$ and expected to influence the negative value of S. After 560 K, the $|S|$ decreases because more hole pockets (bands 6, 7 and so on) are contributing in S. The single band Eqn. 2 confirms that if the more bands of same charge contribute, S will be increased. Therefore, the decrement of magnitude of S at high temperature is directly related to the contribution of multi-band electron and hole pockets. Fig. 4(a) shows that there is a slight deviation between calculated and experimental S at higher temperature range. Here, it is important to note that S is calculated at constant $\mu$, but in general $\mu$ is temperature dependent quantity. Apart from this S is calculated using ground state band-structure which was supposed to be taken as temperature dependent. Therefore, temperature dependent study of $\mu$ and band-structure may improve the result at high temperature region, which is beyond the scope of our present study.

The electrical conductivity per relaxation time ($\sigma/\tau$) is calculated as a function of temperature as shown in Fig. 4(b). Figure shows that $\sigma/\tau$ decreases with temperature initially then increases. This behaviour of $\sigma/\tau$ can be understood by knowing the parameters by which $\sigma/\tau$ is calculated. Here, $\sigma$ of each band is calculated using Eqn. 1. Then total $\sigma$ is computed by adding all the $\sigma$'s corresponding to each band. Temperature dependence of $\sigma$ is taken care by Fermi-Dirac distribution function as appeared in Eqn. 1. Here, it is important to note that $\sigma$ is calculated under constant relaxation time approximation ($\tau_{n}(\varepsilon_{n}(\textbf{k}))=\tau$). Hence, calculated value of $\sigma$ is solely depend on $\textbf{v}_{n}(\textbf{k})$, $-\frac{\partial f}{\partial \varepsilon}$ and number of available states for a given $\mu$. With the increase in temperature, number of states always increases, therefore the initial decrement of $\sigma/\tau$ is directly related to the $\textbf{v}_{n}(\textbf{k})$ in Eqn. 1. At high temperature (above $\sim$520 K), increment of $\sigma/\tau$ is because of rapid increment of number of carriers due to the contribution of mutiple bands (bands 5, 6, 7 and so on in Fig. 3(b)) as similar thing we have seen in the case of S also. From the figure it is clear that the initial decrement of $\sigma/\tau$ is consistent with the experimental trend of $\sigma$ (Fig. 2(b)), but at higher temperature the trend of calculated values are not accordance with the experiment. Here, one can expect that temperature dependent values of $\tau$ may improve the calculated $\sigma$ and give the better match with experiment. Keeping this in mind, the values of $\tau$ are extracted by fitting the calculated $\sigma/\tau$ with experimental $\sigma$. The inset of Fig. 4(b) shows the extracting $\tau$ in the temperature range $300-800$ K. The value of $\tau$ is calculated as $\sim$0.8$\times$10$^{-14}$ s at 300 K, which is in the typical range of 10$^{-14}$ $-$ 10$^{-15}$ s for metals and degenerate semiconductors\cite{ashcroft}. As the temperature increases $\tau$ decreases due to enhancing the number of scattering. Inset of Fig. 4(b) also shows the fitting of $\tau$ with temperature as plotted by solid line. The $\tau$ has the temperature dependency as $\tau\propto1/T^{0.35}$. This behaviour of $\tau$ suggests that calculated $\sigma$ can show decreasing trend on inclusion of temp dependent $\tau$.           

The electronic part of thermal conductivity per relaxation time ($\kappa_{e}/\tau$) is also computed under semi-classical Boltzmann theory. Then $\kappa_{e}$ is calculated by taking temperature dependent $\tau$ (inset of Fig. 4(b)), which is shown in Fig. 4(c). The $\kappa_{e}$ increases with rise in temperature in the whole temperature range. The behaviour of calculated $\kappa_{e}$ can be understood by the following equation\cite{ashcroft}
\begin{equation}
\boldsymbol{\kappa_{e}}=\frac{\pi^{2}}{3}\bigg(\frac{k_{B}}{e}\bigg)^{2}T\boldsymbol{\sigma}
\end{equation}
The temperature dependent variation of $\kappa_{e}$ is decided by T and $\sigma$, since the other symbols of above equation are constants. As $\sigma$ decreases in the full temperaure range, hence T dominates over $\sigma$ for the increment of $\kappa_{e}$. The calculated $\kappa_{e}$ is compared with the experimental $\kappa_{e}$ in the same figure of 4(c). Experimental $\kappa_{e}$ is estimated using Wiedeman-Franz law: $\kappa_{e}=L\sigma T$, where temperature dependent $\sigma$ is taken from experiment (Fig. 2b) and \textit{L} is taken as constant $2.45\times10^{-8}$ $W\Omega/K^{2}$ in the full temperature range. The same value of \textit{L} is previously used by many groups for calculating $\kappa_{e}$ of CoSi\cite{lue,ren,kuo,sun_2017,yu}. The calculated $\kappa_{e}$ are in good agreement with the experimental $\kappa_{e}$ as seen from the figure. But, a slight deviation between experimental and calculated values of $\kappa_{e}$ is observed at higher temperature region (above $\sim$520 K). Therefore, the Wiedeman-Franz law which is derived for the metal\cite{ashcroft} may not be applicable for the fair estimation of $\kappa_{e}$ in such a semimetal CoSi. In addition to this, the \textit{L} is taken as constant which is temperature dependent quantity in general. Hence, temperature dependent \textit{L} may improve the high temperature $\kappa_{e}$ data. The experimental value of $\kappa_{e}$ is found to be $\sim$3.8 W/m-K at 300 K. From this estimation, one can expect that a larger contribution in total $\kappa$ comes from lattice part of $\kappa$ at room temperature. As the temperature increases $\kappa_{e}$ also increases due to the contribution of more energy bands (bands 5, 6, 7 and so on) and hence large number of carriers available at high temperature. The similar kind of contribution of multi-band in the explanation of S and $\sigma$ was also seen as described earlier. At 800 K, the experimental $\kappa_{e}$ is observed as $\sim$7.0 W/m-K, which is almost 50$\%$ of the total $\kappa$.

\begin{figure}
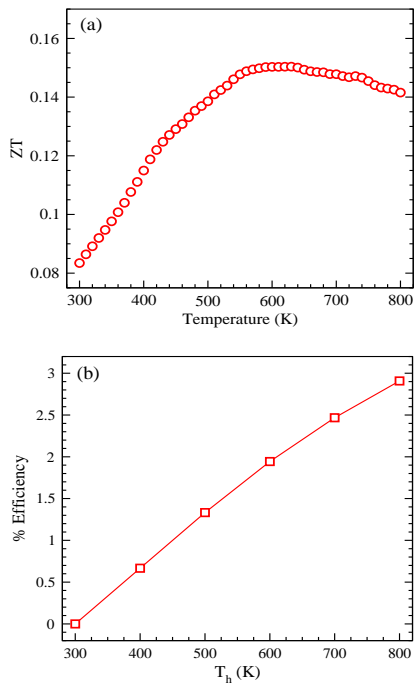
 
\begin{subfigure}{0.36\textwidth}
\includegraphics[width=0.84\linewidth, height=4.3cm]{fig5a_zt.eps} 
\end{subfigure}\\
\vspace{0.3cm}
\begin{subfigure}{0.36\textwidth}
\includegraphics[width=0.835\linewidth, height=4.4cm]{fig5b_efficiency.eps}
\end{subfigure} 
\caption{\small{(a) Temperature dependence of \textit{figure-of-merit}, ZT and (b) $\%$ Efficiency as a function of hot end temperature.}}
\label{fig:image2}
\end{figure}

\subsection{FIGURE OF MERIT AND EFFICIENCY}
The performance of TE materials is characterized by it's \textit{ZT} value and efficiency. In this section, we discuss the \textit{ZT} and efficiency of CoSi. Fig. 5(a) shows the experimental \textit{ZT} values in the temperature range $300-800$ K. The \textit{ZT} at 300 K is found to be $\sim$0.08. As the temperature increases \textit{ZT} is observed to be enhanced rapidly up to $\sim$550 K, mainly due to increment of magnitude of S. Then gives the maximum value of \textit{ZT} $\sim$0.15 at $\sim$630 K. After 630 K, \textit{ZT} decreases slowly till the highest temperature of 800 K with the corresponding values of $\sim$0.14.  

Accurate calculation of efficiency gives the easy way to select the materials in making TEG. Here, we apply the segmentation method to calculate the efficiency of CoSi as proposed by Gaurav \textit{et al}\cite{gaurav}. In this method, efficiency is calculated by varying the hot end temperature, where cold end temperature was kept constant at 300 K. Hot end temperature is varied from 300 to 800 K with step size of 100 K. Efficiency is calculated in every 100 K starting from 300 K by taking every segment temperature ($\Delta T$) as 10 K. The $\%$ efficiency for CoSi is shown in Fig. 5(b) as a function of hot end temperature. Figure shows that efficiency is increasing as the temperature increases. The maximum efficiency is observed as $\sim$3$\%$, when temperatures of cold side and hot side are considered as 300 and 800 K, respectively.   

The measured values of \textit{ZT} and efficiency of CoSi are quite low as compared to conventional TE materials. Although the power factor is high enough, the \textit{ZT} is suppressed by high $\kappa$. Therefore, a rigorous efforts are required to reduce $\kappa$ (which involves $\kappa_{e}$ and $\kappa_{L}$) in order to acquire high \textit{ZT}. Needless to say that the minimizing of $\kappa_{e}$ without affecting $\sigma$ is a difficult job as they involve linear relationship via Wiedeman-Franz law. Hence, the only way to improve \textit{ZT} is by reducing the $\kappa_{L}$ without disturbing the electronic structure of CoSi. The ways like alloying, nanostructuring can be utilized to reduce $\kappa_{L}$\cite{snyder,djsingh}.

\section{CONCLUSIONS}
In this work, we have studied the thermoelectric properties of a novel topological semimetal CoSi by using combined experimental and DFT based methods up to 800 K. For this purpose, we synthesized the CoSi sample using arc melting method and performed the Rietveld refinement which confirms the B20 cubic crystal structure with lattice constants of $a=b=c=4.445$ \AA. First, we have experimentally measured the transport properties viz. S, $\sigma$ and $\kappa$  of CoSi in $300-800$ K range. The S is found to show a non-monotonic behaviour in the studied temperature range with room temperature value of $\sim-$81 $\mu$V/K. The $|S|$ is found to increase up to $\sim$560 K ($\sim-$93 $\mu$V/K) and then decrease till 800 K with a value of $\sim-$84 $\mu$V/K. The measured values suggested the dominating \textit{n}-type behaviour in the full temperature range. The $\sigma$ is found to have monotonic decreasing trend, while $\kappa$ is monotonically increasing in the temperature region studied. The measured value of $\sigma$ ($\kappa$) at 300 K is $\sim$5.2$\times$10$^{5}$ $\Omega^{-1}m^{-1}$ ($\sim$12.1 W/m-K) and reaches the value of $\sim$3.6$\times$10$^{5}$ $\Omega^{-1}m^{-1}$ ($\sim$14.2 W/m-K) at 800 K. The observed temperature dependence of $\kappa$ is found to be of $T^{0.16}$. Further, we have given an insight of measured thermoelectric properties of this topological semimetal up to 800 K, by the combined DFT and Boltzmann transport calculations for electronic properties. The band-structure is calculated by considering SOC which showed the presence of three nodal points. The band-structure analysis suggested that presence of multi-band electron and hole pockets in the dispersion are mainly responsible for deciding the temperature dependent transport behaviour. Specifically, the decrease in magnitude of S above 560 K and change in slope of $\sigma$ around 450 K are found to be due to contribution from the thermally generated charge carriers from the hole pockets. The temperature dependent $\tau$ is calculated by comparing the experimental $\sigma$ and calculated $\sigma /\tau$ which is found to vary with temperature as $\sim$1/$T^{0.35}$. The calculated value of $\kappa_{e}$ using this $\tau$ is found to give fairly good match with experimental $\kappa_{e}$ (obtained from Wiedemann-Franz law). Thus our study suggested that the electronic band-structures obtained from DFT based methods are reasonably good in explaining excited state transport properties of this topological semimetal without the need of beyond DFT methods. However, for better quantitative studies of these properties (specially above $\sim$600 K) we may need to go beyond the conventional DFT methods where temperature dependent band-structures are taken into account.

\section{ACKNOWLEDGMENT}
The authors thank Mr. Gaurav Kumar Shukla and Dr. Sanjay Singh of Indian Institute of Technology (BHU) for the help with sample preparation.

\end{document}